\documentclass[twocolumn,10pt]{wlscirep}
\usepackage{stackengine}

\title{Rhythms of the collective brain: Metastable synchronization and cross-scale interactions in connected multitudes}

\author[1,2,3,*]{Miguel Aguilera}
\affil[1]{Department of Computer Science and Systems Engineering, University of Zaragoza, Zaragoza, 50018, Spain}
\affil[2]{Institute for Cross-Disciplinary Physics and Complex Systems, Palma, 07122, Spain}
\affil[3]{Aragon Institute of Engineering Research, Zaragoza, 50018, Spain}

\affil[*]{sci@maguilera.net}

\begin{abstract}
Crowd behaviour challenges our fundamental understanding of social phenomena.
Involving complex interactions between multiple temporal and spatial scales of activity, its governing mechanisms defy conventional analysis. 
Using 1.5 million Twitter messages from the 15M movement in Spain as an example of multitudinous self-organization, we describe the coordination dynamics of the system measuring phase-locking statistics at different frequencies using wavelet transforms, identifying 8 frequency bands of entrained oscillations between 15 geographical nodes.
Then we apply maximum entropy inference methods to describe Ising models capturing transient synchrony in our data at each frequency band. 
The models show that 1) all frequency bands of the system operate near critical points of their parameter space and 2) while fast frequencies present only a few metastable states displaying all-or-none synchronization, slow frequencies present a diversity of metastable states of partial synchronization.
Furthermore, describing the state at each frequency band using the energy of the corresponding Ising model, we compute transfer entropy to characterize cross-scale interactions between frequency bands, showing 1) a cascade of upward information flows in which each frequency band influences its contiguous slower bands and 2) downward information flows where slow frequencies modulate distant fast frequencies.
\end{abstract}
\begin{document}

\flushbottom
\maketitle
%
%
\thispagestyle{empty}


\section{Introduction}


Coordinated activity is a powerful force in creating and maintaining social ties \cite{mcneill_keeping_2008}. From communal dances in ancient human groups to civic festivals in the French Revolution or goose-stepping formations and stiff right arm salutes in Nazi marches and rallies \cite[p.136, p.148-149]{mcneill_keeping_2008}, visceral, emotional sensations of shared movement have been used to create communal identities and to shape political landscapes.
Historically, forms of distributed communication and coordination have often come together with  episodes of large-scale mobilization and social change, as the widespread print-shop networks of radical reforming movements with the generalization of the printing press during the 16th century German Reformation \cite{hill_anabaptism_2015} or postal networks of the Republic of Letters in the Age of Enlightenment a century later \cite{chang_visualizing_2009}.
Today, amidst unprecedented development of communication technologies, new forms of coordination for large and scattered communities have been unleashed around the globe.

The rise of new digital communication tools and network technologies is accelerating fast bidirectional communication, generating new forms of collective communication and action. Digital communications increase the autonomy and influence of the social groups using them facilitating forms of mass self-communication \cite{castells_communication_2007}, collective intelligence using pools of social knowledge \cite{levy_collective_1999} or \emph{smart mobs} exploiting new found communication and computing capabilities via ubiquitous devices \cite{rheingold_smart_2007}.
From protest movements including the Arab Spring or the Occupy movements to autonomous responses in the face of natural disasters (e.g. Hurricane Sandy or the T\={o}hoku earthquake), several examples highlight the increasing power of digitally connected social and political grassroots movements to shape events.
Recognition of a growing influence has brought with it heightened scholarly interest in its explanation: how such movements arise and self-organize, what mechanisms underlie their formation and how are they able to constitute autonomous social and political subjects?  \cite{bennett_organization_2014}.
Recent advances have described specific elements of connected multitudes: the geographical diffusion of trends \cite{ferrara_traveling_2013}; the interplay between exogenous and endogenous dynamics \cite{oka_self-organization_2014}; or the connection between social media and collective activity in physical spaces \cite{varol_evolution_2014}.
Nevertheless, many of the mechanisms so far explored are specific to a particular scale or level of description of social dynamics. General mechanisms offering explanatory insights across different levels remain poorly articulated.
The same problem applies to qualitative analyses trying to capture general principles of connected multitudes. These include perspectives stressing the individualistic logic pervading digital communication tools operating through sharing personalized content in social media \cite{bennett_logic_2012}, in sharp contrast with narratives highlighting the powerful aggregating and unifying affordances of digital communication tools \cite{juris_reflections_2012}.
We argue that these tensions can be reconciled. Using the analogy of biological brains, what constitutes social collective `brains' as complex entities probably cannot be captured by a single level of description. Instead, it may involve the capacity to display coordination at multiple scales \cite{monterde_multitudinous_2015}, perhaps resembling neural large-scale synchronization over multiple frequency bands \cite{varela_brainweb:_2001,buzsaki_rhythms_2006, le_van_quyen_brainweb_2011}.
Howsoever, the principles operating behind networks of connected multitudes require further conceptual and experimental development to address gaps in extant theory.

Propitiously, the rise of social media and digital data-mining creates the opportunity for a novel analysis of human social systems \cite{lazer_computational_2009} providing mechanisms for explaining their behaviour and opening-up the interactions between different scales of activity for detailed investigation. This opportunity provides an entry point into theoretical debates from where we can begin to generate hypothesis based on inferences from social experimental data. It is a position from which to undertake the difficult task of conceptualizing and describing the interwoven network of causal relations at different levels of description in social systems.

We use a data set of 1.5 million Twitter messages to explore transient phase-locking synchronization as a general mechanism explaining interactions within and between temporal scales.
In particular, we use a well-known social event of large-scale social and political self-organization: the massive political protest of the 15M movement in Spain, emergent in the aftermath of the 2011 Arab Spring and widely thought to be facilitated through digital social platforms \cite{postill_digital_2013}. 
The exemplar of the 15M movement is interesting for a number of reasons: First, it consists in a self-organized social movement arising from online communication in a distributed network of citizens and civil associations (without significant coverage by mainstream media until days after the protests had taken to the streets). 
Second, the movement led to massive, nationwide demonstrations and encampments, creating a decentralized collective agency which has had a profound impact in Spanish politics \cite{sampedro_spanish_2014, pena-lopez_spanish_2014, tormey_reinventing_2015}.
Finally, a series of studies have characterized some of the emerging properties of the 15M and how it exhibits features typical of critical systems and distributed self-organization \cite{borge-holthoefer_structural_2011, borge-holthoefer_dynamics_2016}.

Using this data set, we propose phase-locking statistics between geographical nodes  at different frequencies  as a generic description of coordination in a nationwide social system. This description allows us to use maximum entropy techniques to extract Ising models mapping the statistical mechanics of the system at each frequency band and thus obtain a deeper understanding of the spatiotemporal patterns of coordination within and between frequency bands.
Inspecting the properties of the models at each frequency band we observe that all bands are operating near a critical point but that different frequencies play different roles in the system. While fast bands alternate states of (almost) full synchronization and full desynchronization, bands with slower frequencies display a wide range of possible configurations of metastable states with clusters of partial synchronization.
Furthermore, applying transfer entropy in the energy landscape at each frequency described by the Ising models, we characterize cross-scale interactions showing an asymmetry between upward and downward influences, where high frequency synchronization influences nearby slower frequencies, while slow frequency bands are able to modulate distant faster bands.
We argue that our results offer a promising step towards the description of general mechanisms operating at different scales, suggesting the existence of general rules for scaling up and down the dynamics of multitudinous collective systems.

\section{Results}

We use a data set of 1,444,051 time-stamped tweets from 181,146 users, collected through the Twitter streaming API between 13 May 2011 and 31 May 2011 \cite{pena-lopez_spanish_2014} using T-Hoarder \cite{congosto_t-hoarder:_2017}. 
Messages were captured during 17 days during the Spanish 15M social unrest events in 2011 containing at least one of a set of 12 keywords or hashtags related to the protest (see reference \cite{pena-lopez_spanish_2014} for a detailed description).
We extracted geographical information from the location information of users (see Supplementary Information), selecting the 15 urban areas with the largest number of messages. Using this information, we generated time-stamped series reflecting the number of tweets emitted from each city for intervals of 60 seconds.

\subsection{Synchronization at multiple frequencies}

One of the most prominent features of the $15$M movement was its fast territorial development. Without any coordination centre or any formal organization, the movement was able to reproduce a network of camps across Spanish cities in a period of a few days. As this coordination between geographical nodes takes place at several temporal scales, we propose a generic description of these interactions based on the temporal coordination of oscillations at multiple frequencies.
We analyse the coordination between populations at main Spanish cities using Morlet wavelet filtering to extract the phase content $\theta_i(f,t)$ of the activity time series at city $i$  at time $t$ and frequency $f$, with a span of frequencies in the range [$1.67\cdot 10^{-3} Hz$, $9.26\cdot 10^{-5} Hz$] (from 10 minutes to 3 hours) logarithmically distributed with intervals of $10^{0.01}$. 
We use phase-locking statistics \cite{lachaux_studying_2000} to define phase-locking values between two cities $i$ and $j$ as:

\begin{equation}
  \phi_{ij}(f,t)= \biggl\lvert { \frac{1}{\delta} \sum\limits_{\tau =-\delta/2}^{ \delta /2}{e^{ \mathrm{i} (\theta_{y}(f,t+\tau)-\theta_{x}(f,t+\tau))}} }  \biggl\rvert \cdot A_{ij}(t)
  \label{eq:plv}
\end{equation}
where $\delta$ is the size of the window of temporal integration: $\delta=\frac{n_{c}}{f}$, being $n_{c}$ the number of cycles in which we analyse phase-locking. We use a value of $n_{c}=8$ cycles, similar to the values typically used in neuroscience, ensuring that we are detecting sustained synchronization. $A_{ij}(t)$ is a corrector factor removing spurious synchronization when the network is inactive (e.g. during nighttime, see Supplementary Information).

Statistical significance of phase-locking values is determined by comparing them to phase-locking values of surrogate time series obtained using the amplitude adjusted Fourier transform \cite{schreiber_improved_1996}. We use $200$ surrogate time series to estimate a significance threshold for the values of $\phi_{ij}(f,t)$ for all values of $f$. The average phase-locking values of surrogate time series were used to compute a threshold $\phi_{th}(f)$, indicating a value higher than $99\%$ of surrogate data.
Using this threshold, we define  phase-locking links between two cities $i$ and $j$ as statistically salient values of $\phi_{ij}(f,t)$:
\begin{equation}
  \Phi_{ij}(f,t) =
      \begin{cases}
      1, & \text{if}\ \phi_{ij}(f,t) \geq \phi_{th}(f)  \\
      0, & \text{otherwise}
    \end{cases}
  \label{eq:pll}
\end{equation}

As we document in \cite{aguilera_collective_2015}, using phase-locking statistics we find widespread moments of significant synchrony at different instants often corresponding with important moments of the 15M protests. 
As well, in the Supplementary Information (Section S4) we provide an analysis of the stability of the synchronization patterns found by wavelet filtering in comparison with other choices of window width for filtering the data.

\begin{figure}
\centering
\includegraphics[width=8cm]{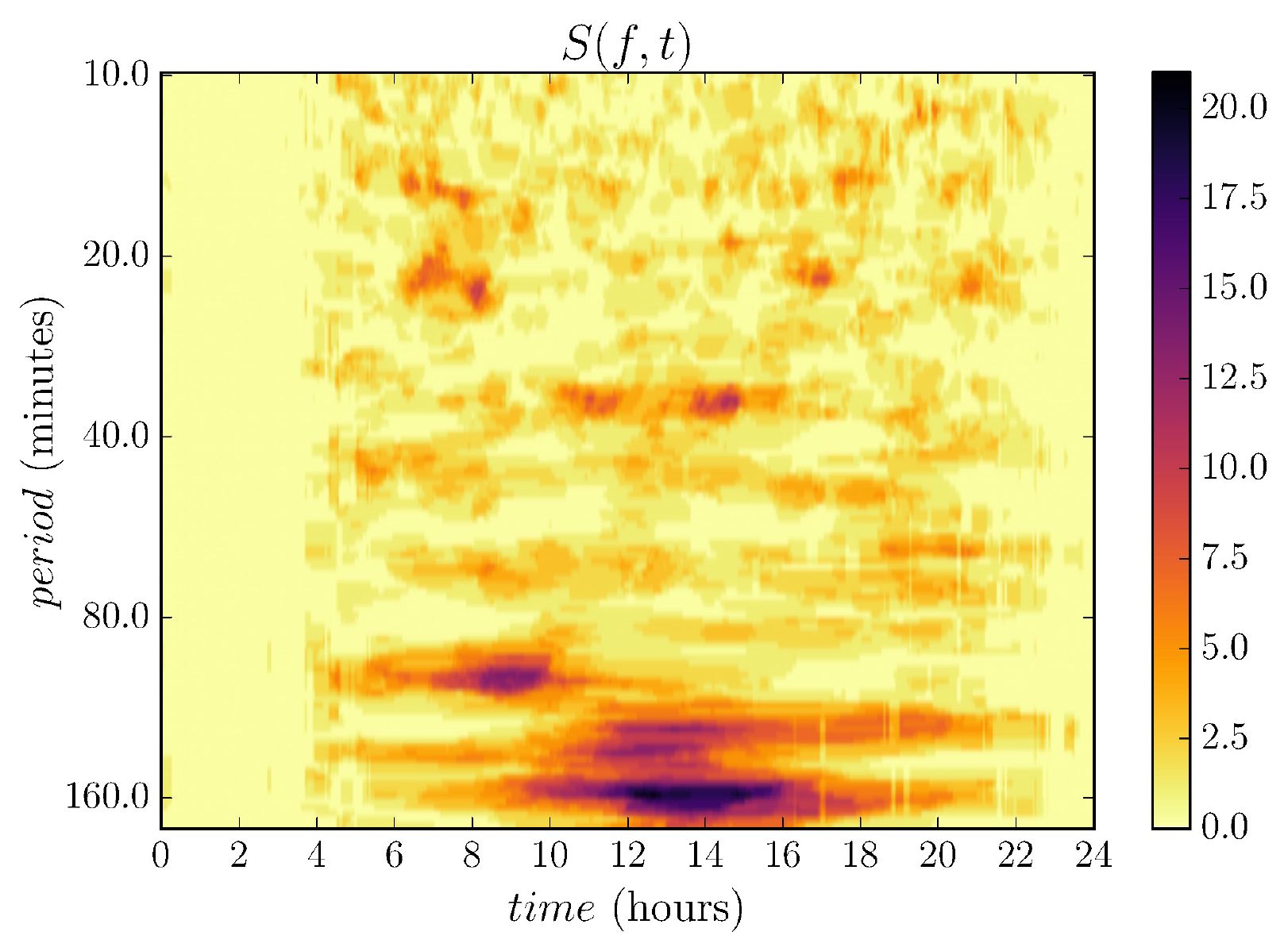}
\caption{\textbf{Phase-locking statistics.} Sum of the number of phase-locking links between all cities  $S(f,t)$ for day May 27. The horizontal axis represents a temporal span of instants $t$ during the day, and the vertical axis represents the period corresponding to the frequency $f$ of the wavelets used for extracting phase contents of the signals. The colour represents the sum of phase-locking links for $f$ and $t$.}
\label{fig:PLS}
\end{figure}

For illustrative purposes, in  Figure \ref{fig:PLS} we show the total number of phase-locking links $S(f,t)=\sum_{i,j} \Phi_{ij}(f,t)$ for a specific day of the protests.
At faster frequencies (lower period) we observe short and less intense instants of synchrony, while at slower frequencies synchrony lasts for longer periods of time.
Using wavelet pattern matching \cite{du_improved_2006} over $S(f)=\langle S(f,t) \rangle$ after applying a linear detrending, we detect frequency peaks of synchronization in the system (see Supplementary Information, Figure S1, Table S1), identifying eight main frequency bands of synchronization $f_k$,  $k=1, ..., 8$, where larger $k$ corresponds to larger timescales (i.e. slower frequencies). 

\subsection{Pairwise maximum entropy modelling of phase-locking statistics}

In order to inspect how these phase-locked coalitions are operating at each frequency band, we derive from our data statistical mechanics models of the system. With these models we can infer macroscopic properties from microscopic descriptions of the system.
Specifically, we use Ising models, which consist of discrete variables that traditionally are assumed to represent the magnetic moments of atomic spins that can be in one of two states (+1 or -1). In our case positive spins will represent the presence of synchronizing activity of a node at a particular frequency. Spins are connected to other spins in the networks, allowing pairwise interaction between nodes.  
This is the least-structured (i.e. maximum entropy) model that is consistent with the mean activation rate and correlations of the nodes in the network. Pairwise maximum entropy models have been successfully used to map the activity of networks of neurons \cite{schneidman_weak_2006}, antibody sequences \cite{mora_maximum_2010} or flocks of birds \cite{bialek_statistical_2012}. These models, instead of being postulated as approximations of real phenomena, can infer exact mappings capturing measured properties of a system (means and correlations in our case), making them good candidates for capturing the structures underlying social coordination.

Using Ising models we infer the probability distribution of possible states $s$ of the network at a specific synchronization frequency, corresponding to all the combinations of binary possibilities of each node being or not being phase-locked to other nodes in the network. For simplicity, we consider the state of a node $i$ equal to one when the node is active in a synchronized cluster (i.e. that is $s_i=1$ when $\max\limits_j [\Phi_{ij}]=1$), and otherwise the state of the node is set to  $s_i=-1$.

The maximum entropy distribution consistent with a known average energy is the Boltzmann distribution $P(s) =   Z^{-1} e^{-\beta E(s)}$, where $s$ is a state of the network, $Z$ is the partition function and $\beta = \frac{1}{T k_B}$, being $k_B$ Boltzmann's constant and $T$ the temperature. The energy of the model with pairwise interactions is defined as $E(s) = - \sum_i h_i s_i - \frac{1}{2}  \sum_{i < j} J_{ij} s_i s_j$, where `magnetic fields' $h_i$ represent influences in the activation of individual nodes  and `exchange couplings' $J_{ij}$ stand for the tendencies correlating the activity between nodes.
Without loss of generality we can set the temperature $T = 1$. Considering a pairwise model, the resulting distribution of the maximum entropy model is:

\begin{equation}
 P(s) =   \frac{1}{Z} exp \Bigg[ {\beta \sum_i h_i s_i + \frac{1}{2} \sum_{i < j} J_{ij} s_i s_j}\Bigg]
 \label{eq:Ising}
\end{equation}
where the $h_i$ and $J_{ij}$ are adjusted to reproduce the measured mean and correlation values between nodes in the network.

From the frequency bands $f_k$ extracted in the previous section we extract models of pairwise correlations at the corresponding frequencies. 
For each frequency band, we infer an Ising model $P_{f_k}(s)$ solving the corresponding inverse Ising problem, using a coordinate descent algorithm (see Methods) for fitting the parameters $h_i$ and $J_{ij}$ that reproduce the means and correlations found in the series of states $s$ for the description of phase-locking relations at each frequency. 

The accuracy of the inferred models can be evaluated by testing how much of the correlation structure of the data is captured. One measure to evaluate this is the ratio of multi-information between model and real data \cite{mora_are_2011}. In our case, our data limits us to computing the entropy of small sets of nodes (between 5 and 7). Limiting our entropy calculations to random sets of five to seven nodes (see Table S2), we can see in Figure S3 and Table S3 that our  models are able to capture around 70\% of the correlations in the data for subsets of the indicated sizes (see Supplementary Information for a detailed description).

\begin{figure*}
 \centering
 \begin{tabular}{lll}
  \topinset{\bfseries(A)}{\includegraphics[height=50mm,keepaspectratio=true]{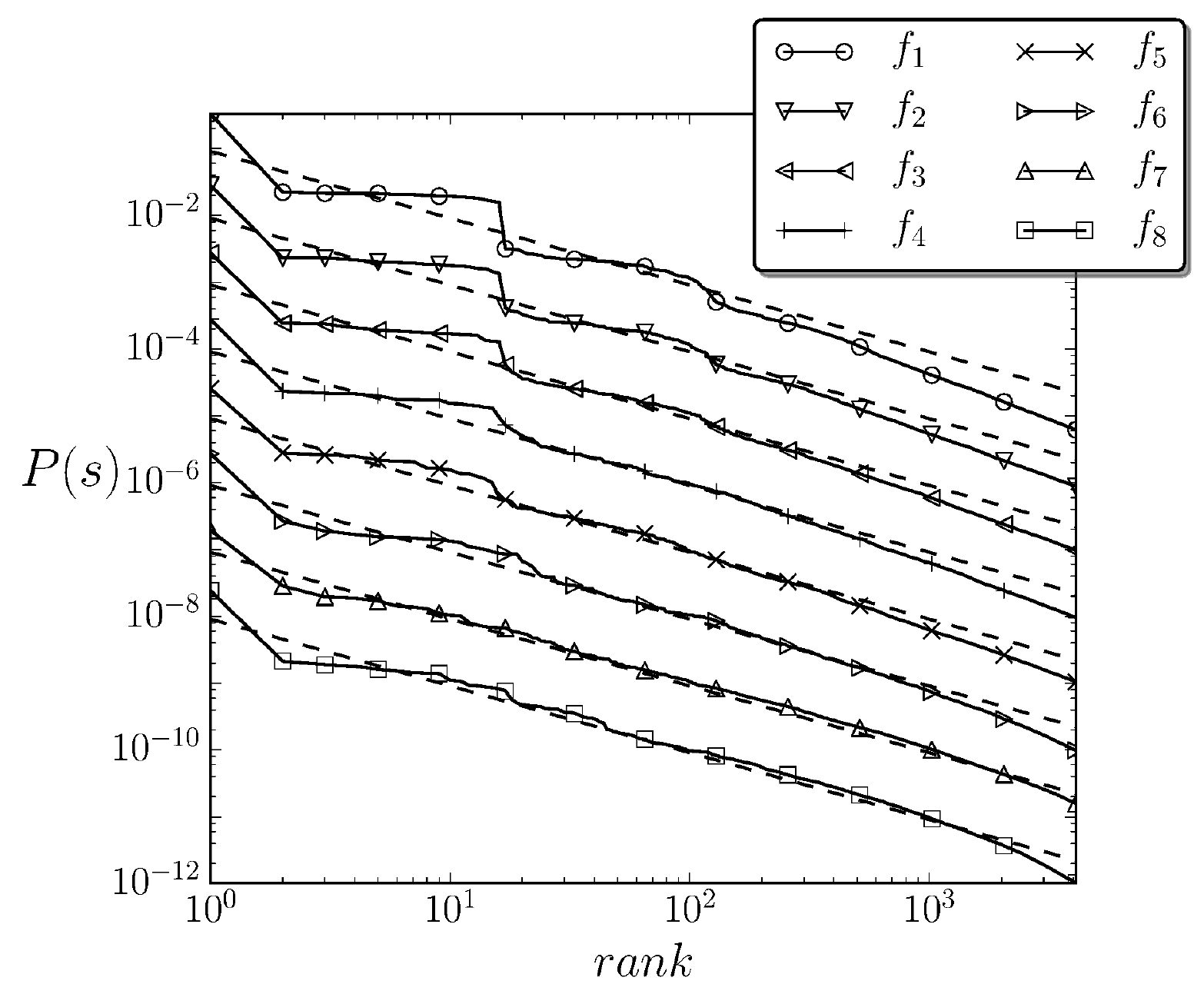}}{0.0in}{-1.0in} &
  \topinset{\bfseries(B)}{\includegraphics[height=50mm,keepaspectratio=true]{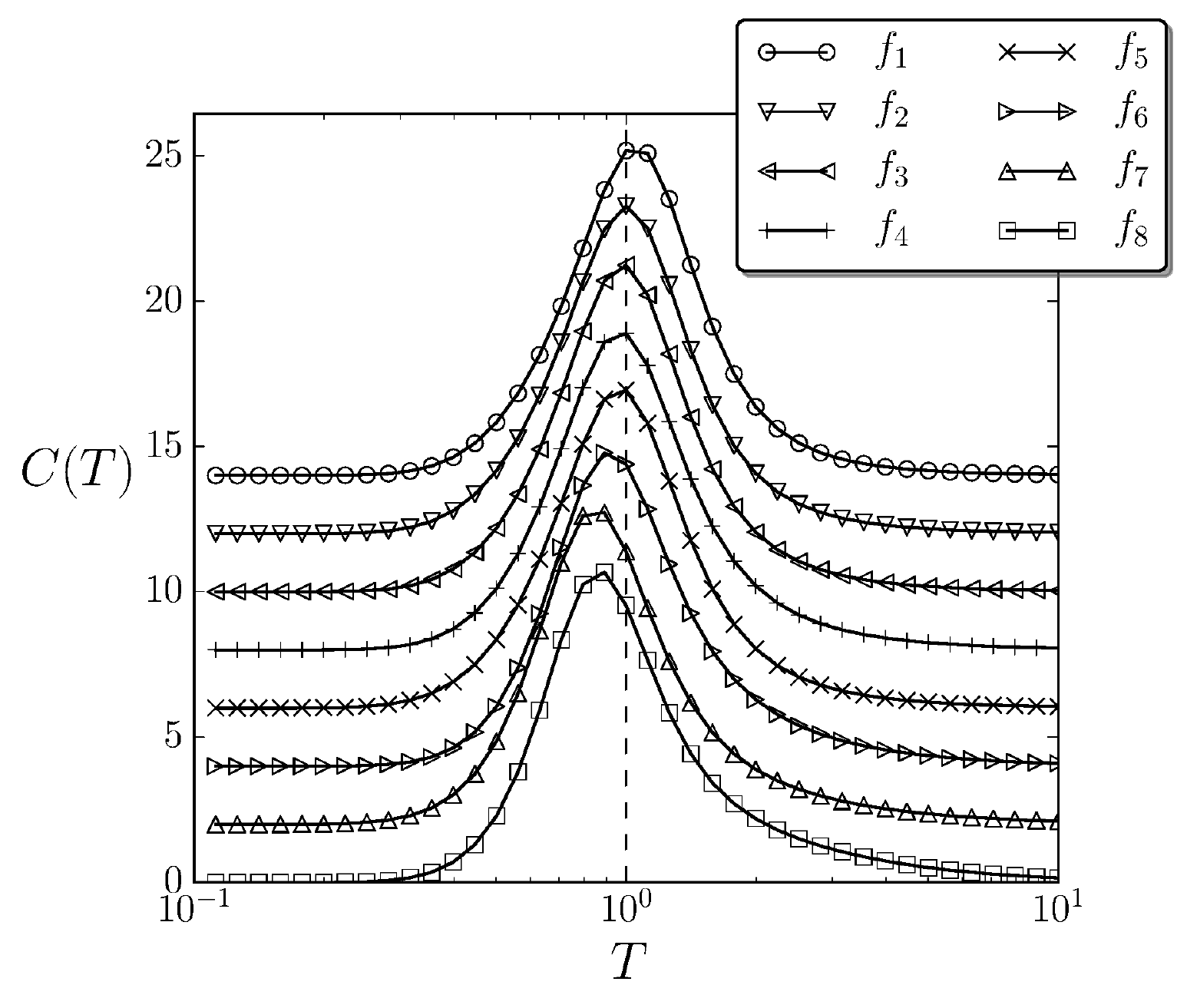}}{0.0in}{-1.0in} &
  \topinset{\bfseries(C)}{\includegraphics[height=50mm,keepaspectratio=true]{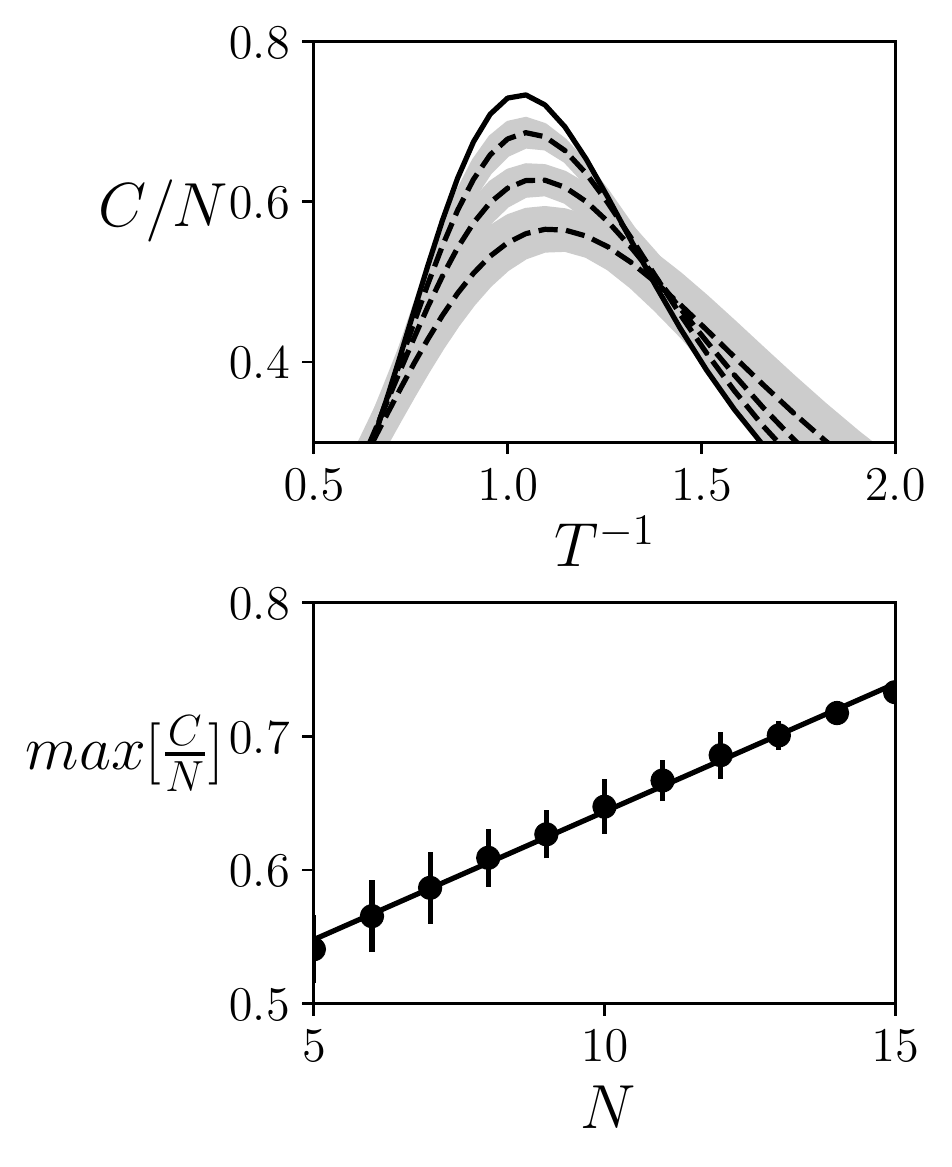}}{0.0in}{-0.7in}
 \end{tabular}
  \caption{\textbf{Signatures of criticality}.  
  \textbf{(A)} Ranked probability distribution function of the inferred Ising models for the different frequency bands (solid lines) versus a distribution following Zipf's law, (i.e. $P(s) = 1/rank$, dashed lines).
  \textbf{(B)} Heat capacity versus temperature for the inferred Ising models for different frequency bands.  Temperature $T=1$ is the point where our models are poised, coinciding with a divergence of the heat capacity. Lines are shifted vertically in order to facilitate visualization. 
  \textbf{(C)} Divergence of the normalized heat capacity with the size of the system, showing (top) the peaks of the normalized heat capacity averaged over 100 Ising models at $f_5$ for sizes 6,9,12 (dashed lines) and 15 (solid line) and (bottom) linear trend of the peak maxima of the averaged heat capacity for sizes 5 to 15. Grey areas represent the error bars.}
 \label{fig:criticality}
\end{figure*}

Once we have extracted a battery of models $P_{f_k}(s)$, indicating the probability distributions of phase-locking configurations at different frequency bands, we explore the thermodynamic (macroscopic) properties associated to them. First, we observe that all the models are poised close to critical points. One signature of criticality we find is that the probability distribution of $P_{f_k}(s)$ follows a Zipf's law (Figure \ref{fig:criticality}.A), specially for slower values of $f_k$. Finding a scale-free distribution in our model is consistent with power laws appearing in the dynamics of the temporal series of tweet activity found in this data set \cite{aguilera_collective_2015} or in structural parameters in similar data sets \cite{borge-holthoefer_structural_2011}. 
Nevertheless, the sole occurrence of a power law is generally insufficient to assess the presence of criticality and may arise naturally in some non-equilibrium conditions. Thus, further evidence is necessary to test if the system is in a critical point.

The Ising model allows us to find further evidence of the critical behaviour of the model by exploring divergences of some variables in its parameter space. 
By introducing a fictitious temperature parameter $T$  (previously assumed to be equal to 1), we can explore the parameter space of the system and look for critical points.
Modifying the value of $T$ is equivalent to a global rescaling of the parameters of the agent transforming $h_{i} \leftarrow  h_{i}/T$ and $J_{ij} \leftarrow J_{ij}/T$, thus exploring the parameter space along one specific direction.

Specifically, a sufficient condition for describing a critical point in the parameter space of an Ising model is the divergence of its heat capacity, which is defined as:
\begin{equation}
 C(T)=T \frac{\partial H[P(s)]}{\partial T} = \frac{1}{T^2} \langle E^2(s) \rangle - \langle E(s) \rangle ^2
\end{equation}
where $H[P(s)]$ is the Shannon Entropy of the probability distribution of an Ising model. 
A divergence in the heat capacity of the system is an indicator of critical phenomena. As we observe in Figure \ref{fig:criticality}.B, for all $f_k$ the peak of the heat capacity is around the value $T=1$, suggesting that the  models are poised near critical points. Inferring the Ising models to match random subsets of the network nodes (see Supplementary Information), we observe how the normalized peak in the heat capacity averaged over 100 random models diverges with the system size (the specific -- although representative -- case of $f_5$ is shown in Figure \ref{fig:criticality}.C, see Figure S5 for other frequencies), where $C(T)/N$ grows with $N$ with a nearly linear rate in the range $[0.0188,0.0207]$ (see Supplementary Information). Together with the Zipf distribution, the divergence of the heat capacity suggests that social coordination phenomena in the 15M social network are operating in a state of criticality \cite{mora_are_2011}.

The fact that all frequency bands are operating near critical points does not mean that they are displaying the same behaviour.
We can extract more information about the behaviour of the system at each frequency by analysing the presence of locally-stable or metastable states in the system. Metastable states are defined as states whose energy is lower than any of its adjacent states, where adjacency is defined by single spin flips. This means that in a deterministic state (i.e. a Hopfield network with $T=0$) these points would act as attractors of the system. In our statistical model metastable states are points in which the system tends to be poised, since their probability is higher than any of its adjacent states. Finding the metastable states of the models at each frequency, we observe how the number of metastable states increases for slower frequencies (Figure S4.B), as the model presents a higher number of negative (inhibitory) couplings $J_{ij}$ (see Figures S4.A, S6.B). A detailed list of metastable states and their basins of attraction can be found in Table S4.

\begin{figure}
\centering
\includegraphics[width=8cm]{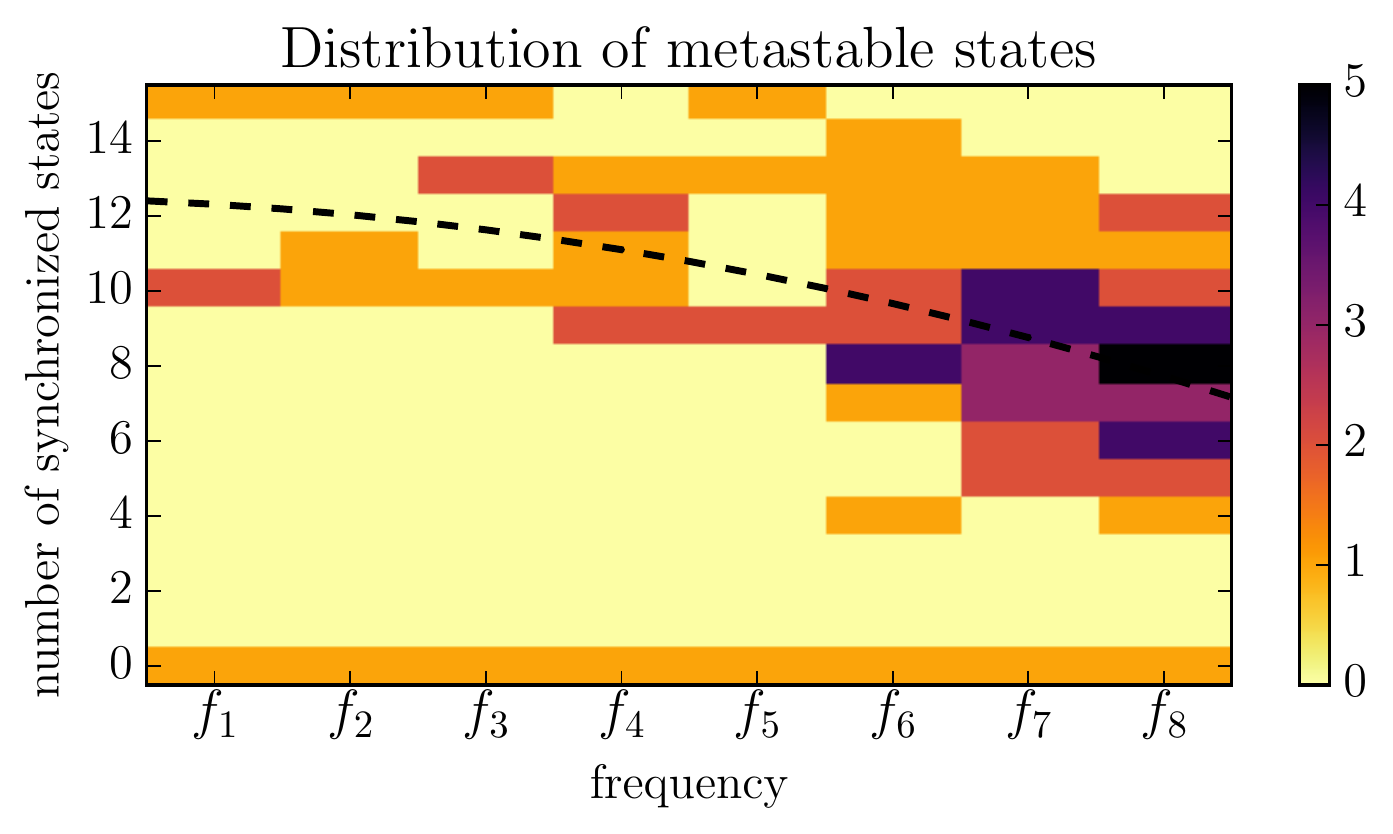}
\caption{\textbf{Distribution of metastable states by frequency and synchronized nodes}. Distribution of the number of active (i.e. synchronized) states of the inferred Ising models for each frequency band. The horizontal axis represents the frequency band, the vertical axis the number of active states of the metastable states and the color represents the count of metastable states with the same number of active states. The dashed line is the result of a least squares second order polynomial fit (higher orders show a similar trend) over the number of active nodes of metastable states (excluding states with zero active nodes) with respect to the frequency index, showing a decrease of the number of synchronized nodes in the metastable states in favour of local clusters  of activity.}
\label{fig:metastable-states}
\end{figure}

Moreover, if we count the number of nodes that are phase-locked (i.e. the sum of all nodes with $s_i=1$) for each metastable state represented in Figure \ref{fig:metastable-states}, we observe important distinctions among frequency bands. For faster values of $f_k$ there are only a few metastable states: a state where all nodes are not phase-locked (i.e. the system is completely desynchronized), and a few values where almost all nodes are phase-locked. Thus, at fast frequencies synchronization rapidly spreads from zero to all nodes in the network. On the other hand, for slower frequencies the number of metastable states grows and the number of phase-locked nodes for each state decreases. This shows that slow frequency synchronizations allows the creation of a variety of clusters of partial synchronization, allowing parts of the network to sustain a differentiated behaviour.

These results suggest that fast and slow synchronization frequencies in the network operate in complementary regimes --all operating near critical points-- the former rapidly propagating information to all the network and the latter sustaining a variety of configurations responding to specific situations.
Systems in critical points present a wide range of dynamic scales of activity and maximal sensitivity to external fluctuations. These features may be crucial for large systems that are self-organized in a distributed fashion.
The presence of these complementary modes of critical behaviour at different frequency bands suggests that the system might be operating in a state of self-organized criticality, in which frequency bands adaptively regulate each other in order to maintain a global critical behaviour.

\subsection{Cross-scale interactions between frequency bands}

Modelling phase-locking statistics provides a characterization of the interactions within frequency bands of synchronization. Furthermore, differences in the metastable states at each frequency band suggest what kind of interactions take place between distinct temporal scales. Because our definition of phase-locking statistics is restricted to interactions within the same frequency, we cannot use the computed phase-locking statistics to directly model inter-scale phase-locking between different frequencies (e.g. 2:1 phase-locking). However, we can use the thermodynamic descriptions of the system provided by maximum entropy models to simplify the analysis of inter-scale relations in real data.

Analysis of multiscale causal relations is typically a difficult task, and in our case we have to deal with a system of a high number of dimensions ($15\cdot8=120$ dimensions). Nevertheless, the Ising models describe the stability of the configurations of the 15 nodes in the network at each frequency band with an energy value. Thus, an easier way to describe multiscale interactions is to observe how fluctuations in the energy at one level affect the energy of the system at other levels, reducing the dimensions we have to deal with to only the $8$ frequencies of synchronization. 

\begin{figure*}
\centering
  \topinset{\bfseries(A)}{\includegraphics[height=5cm]{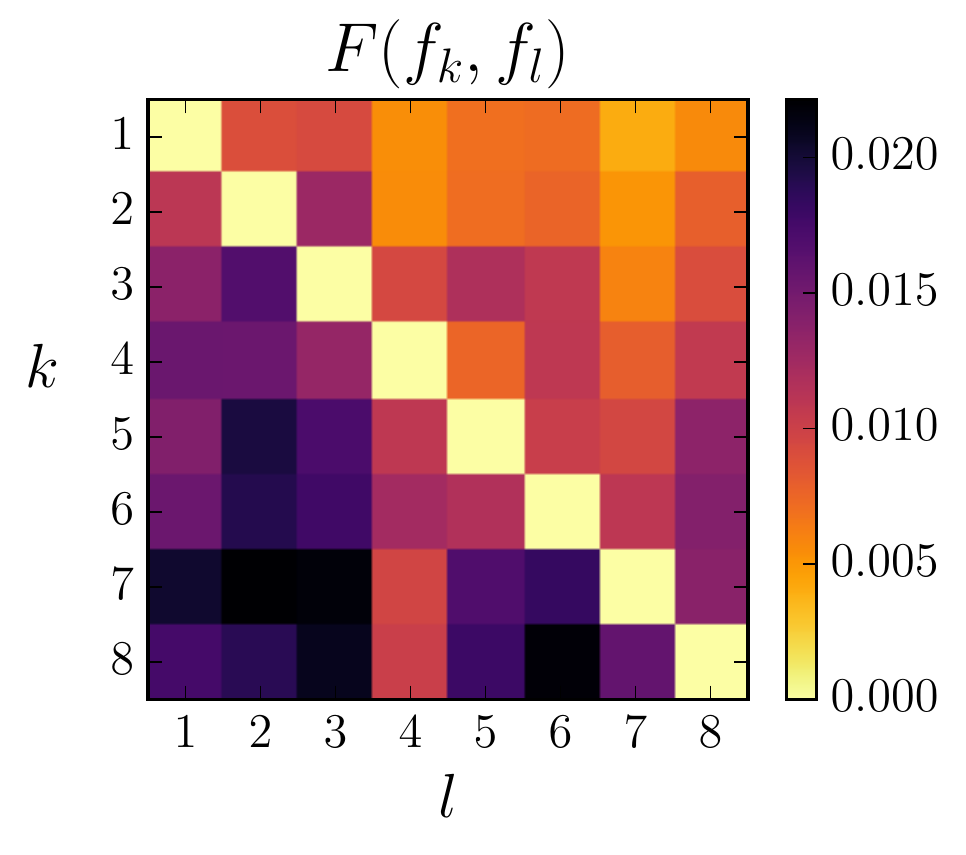}}{0.in}{-0.9in}
  \topinset{\bfseries(B)}{\includegraphics[height=5cm]{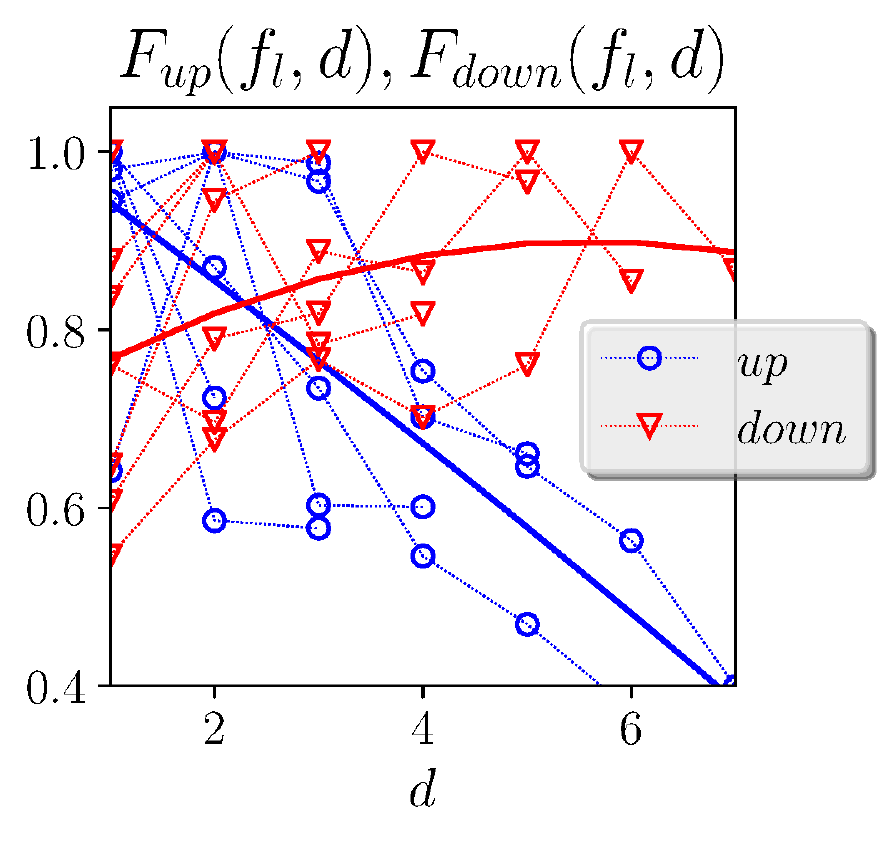}}{0.in}{-0.9in}
\caption{\textbf{Average transfer entropy}. \textbf{(A)} Average transfer entropy $F(f_k,f_l)$ across values of $\tau$ between energy levels at different frequencies. \textbf{(B)} Tendencies of upward ($F_{up}(f_l,d)$, blue circle markers) and downward ($F_{down}(f_l,d)$, red triangular markers) transfer entropy values respect to the distance $d$ between frequencies. Solid lines represent a least squares second order polynomial fit (higher orders show a similar trend) of the values of $F_{up}(f_l,d)$ (blue solid line) and $F_{down}(f_k,d)$ (red solid line) respect to $d$.  We observe how upward transfer entropy values display an important decrease with distance, whereas downward transfer entropy presents slightly larger values for larger distances.}
\label{fig:TEmean}
\end{figure*}

We characterize the information flow between frequency bands using transfer entropy  \cite{schreiber_measuring_2000} between energy levels at each frequency $E_{f_k}(s)$. Transfer entropy captures the decrease of uncertainty in the state of a variable $Y$ derived from the past state of other variable $X$:

\begin{equation}
\begin{split}
 \mathcal{T}_{X \rightarrow Y}(\tau) = \sum\limits_{x_{t+\tau},x_{t},y_{t}} P(x_{t+\tau},x_{t},y_{t}) log \frac{P(x_{t+\tau}|x_{t},y_{t})}{P(x_{t+\tau}|x_{t})}
 \end{split}
 \label{eq:TE}
\end{equation}
where $x_{t}$ denotes the state of $X$ at time $t$ and $\tau$ indicates the temporal distance used to capture interactions.

In order to compute transfer entropy over energy values between timescales, we discretize the values of energy $E_{f_k}(s)$ into a variable with 3 discrete bins $E^*_{f_k}(s)$ using the Jenks-Caspall algorithm \cite{jenks_error_1971}. The value of 3 bins was selected to optimize the computation of joint probability density functions (see Supplementary Information) although we tested values from 2 to 6 bins with similar results.
Using transfer entropy we estimate the causal interactions between energetic states at each timescale by computing the values of $\mathcal{T}_{E^*_{f_k}\rightarrow E^*_{f_l}}(\tau)$ (see Figure S7 for a representation of transfer entropy functions) for values of $\tau$ between $1$ and $2^9$ minutes  (i.e. up to 8.5 hours) logarithmically distributed with intervals of $2^{0.25}$.

\begin{figure}[ht]
\centering
\includegraphics[width=8cm]{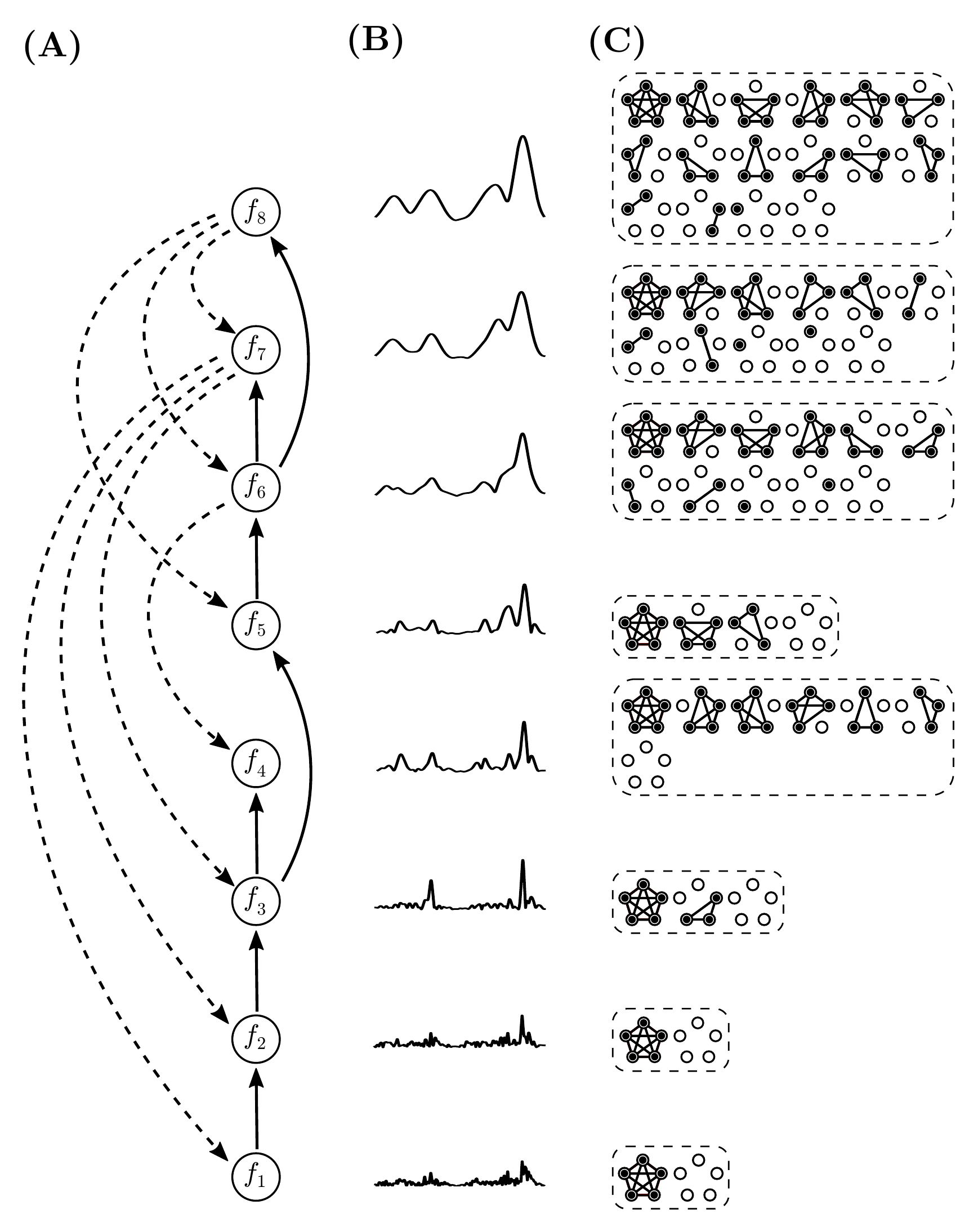}
\caption{\textbf{Cross-scale interactions in social coordination}. Schematic displaying the results presented in previous figures. \textbf{(A)} Interactions in terms of average transfer entropy between energy levels at different frequencies described by $F_{up}(f_l,d)$ and $F_{down}(f_l,d)$. For simplicity only the largest upward (solid lines) and downward (dashed lines) values of transfer entropy are represented. \textbf{(B)} Fragment of activity of all nodes in the network (tweets per second) filtered by the 8 wavelets used in the filtering. \textbf{(C)} Metastable states of the models at each frequency. For simplicity, only metastable configurations of the five cities with largest volume of activity are represented.}
\label{fig:cross-scale}
\end{figure}

To simplify the interpretation of the data, we compute the average value of transfer entropy (across the logarithmic range of $\tau$) for pairs of frequencies as $F(f_k,f_l) = \langle \mathcal{T}_{E^*_{f_k}\rightarrow E^*_{f_l}}(\tau) \rangle$ (Figure \ref{fig:TEmean}.A). Moreover, we separate the values of upward and downward flows of information for each node, characterizing $F_{up}(f_l,d) = F(f_{l-d},f_{l}) / \max_d [F(f_{l-d},f_{l})]$ and $F_{down}(f_l,d) = F(f_{l+d},f_{l}) / \max_d [F(f_{k+d},f_{k})]$, where $d$ takes values between $1$ and $7$, and upward and downward entropies are divided by their maximum values in order to compare transfer entropy between nodes with distinct values of entropy.
In Figure \ref{fig:TEmean}.B, we observe upward and downward flows of information. As we can see, upward flows decrease importantly with distance between scales. In contrast, downward flows increase slightly with distance between scales.

These results show an interesting picture of cross-scale interactions. While in upward interactions energy at each frequency band only influences neighbouring slower bands, in downward interactions slow frequency bands modulate distant faster bands. We also observe this in the schematic in Figure \ref{fig:cross-scale}.A, where for simplicity only the largest values of $F_{up}(f_l,d)$ and $F_{down}(f_l,d)$ are displayed for each frequency band. 
These results suggest that there might be general rules for scaling up and scaling down social coordination dynamics in a nested structure of frequency bands. The mechanisms involved might resemble those found in neuroscience, where upward cascades have been found to take place in the form of avalanches propagating local synchrony and downward cascades take the form of phase-amplitude modulation of local high-frequency oscillations by large-scale slow oscillations \cite{le_van_quyen_brainweb_2011}.
Future research is required for testing the application of these rules to other social coordination phenomena and the specific mechanisms operating behind upward and downward cross-scale interactions.

\section{Discussion}


It is appealing to think that general coordinative mechanisms may be suited to explain the behaviour of social systems at different scales.
Here, using a large-scale social media data set, we have shown how the application of maximum entropy inference methods over phase-locking statistics at different frequencies offers the prospect of understanding collective phenomena at a deeper level. 
The presented results provide interesting insights about the self-organization of digitally connected multitudes. Our contribution shows that phase-locking mechanisms at different frequencies operate in a state of criticality for rapidly integrating the activity of the network at fast frequencies while building-up an increasing diversity of distinct configurations at slower frequencies.
Moreover, the asymmetry between upward and downward flows of information suggests how social systems operating through distributed transient synchronization may create a hierarchical structure of temporal timescales, in which hierarchy is not reflected in a centralized control but in the asymmetry of information flows between the coordinative structures at different frequencies of activity. This offers a tentative explanation of how an unified collective agency, such as the 15M movement, might emerge in a distributed manner from mechanisms of transient large-scale synchronization. 
Of particular interest would be to test the extent our findings about the structural and functional relations of social coordination apply to other self-organizing social systems, or their relation with mechanisms of cross-scale interactions known from large-scale systems neuroscience. 
A new generation of experimental findings based on statistical mechanics models may provide the opportunity to discover the mechanisms behind multitudinous social self-organization.

\section*{Methods}


\subsection*{Data availability}
The data employed in this study was kindly provided by the authors of \cite{pena-lopez_spanish_2014}. 

\subsection*{Learning pairwise maximum entropy models from data}
Ising models are inferred using an adapted version of the coordinate descent algorithm described in \cite{dudik_performance_2004}. The coordinate descent algorithm works by iteratively adjusting a single weight $h_i$ or $J_{ij}$ that will maximize an approximation of the change in the empirical logarithmic loss between the observed data and the model, computed through the means and correlations present in the empirical data and the model.
The code implementing the coordinate descent algorithm is available at \url{https://github.com/MiguelAguilera/ising}.


\section*{Acknowledgements}


Research was supported in part by the Spanish National Programme for Fostering Excellence in Scientific and Technical Research project PSI2014-62092-EXP and projects TIN2016-80347-R and FFI2014-52173-P funded by the Spanish Ministry of Economy and Competitiveness.

%

\section*{Competing interests}

The author declares no competing financial interests.

\section*{Supplementary Materials}

The Supplementary Information file provides additional detailed information about the techniques and results of the contributions of this article.

\bibliography{references}

\end{document}